\documentclass[aps,prb,twocolumn,showpacs,superscriptaddress]{revtex4-1}

\usepackage{graphicx}
\usepackage{cancel}
\usepackage{amsmath, amsthm, amsfonts}
\usepackage{xcolor}
\usepackage{bm}
\usepackage{multirow}

\newcommand{\beq}{\begin{equation}}  
\newcommand{\eeq}{\end{equation}}  
\newcommand{\beqa}{\begin{eqnarray}}  
\newcommand{\eeqa}{\end{eqnarray}}

\newcommand{\expESR}{Baumann_Paul_science_2015,Natterer_Yang_nature_2017,Choi_Paul_natnano_2017,
Willke_Paul_sciadv_2018,Willke_Bae_science_2018,Yang_Bae_prl_2018,Willke_Yang_arXiv_2018,
Yang_Willke_natnano_2018}

\begin{document}

\title{A cotunneling mechanism for all-electrical Electron Spin Resonance of single adsorbed atoms}

\author{J. Reina G\'alvez}
\affiliation{Centro de F\'{\i}sica de Materiales CFM/MPC (CSIC-UPV/EHU), 
Paseo Manuel de Lardiz\'abal 5, 20018 Donostia-San Sebasti\'an, Spain}

\author{C. Wolf}
\affiliation{Department of Physics, Ewha Womans University, Seoul 03760, Republic of Korea}
\affiliation{Center for Quantum Nanoscience, Institute for Basic Science, Seoul 03760, Republic of Korea.}

\author{F. Delgado}
\email[E-mail: ]{fernando.delgado@ull.edu.es}
\affiliation{Instituto de estudios avanzados IUDEA,
 Departamento de F\'{i}sica, Universidad de La Laguna, C/Astrof\'{i}sico Francisco S\'anchez, s/n. 38203, Tenerife, Spain}
\affiliation{Donostia International Physics Center, 
Paseo Manuel de Lardiz\'abal 4, 20018 Donostia-San Sebasti\'an, Spain}

\author{N. Lorente}
\affiliation{Centro de F\'{\i}sica de Materiales CFM/MPC (CSIC-UPV/EHU), 
Paseo Manuel de Lardiz\'abal 5, 20018 Donostia-San Sebasti\'an, Spain}
\affiliation{Donostia International Physics Center, 
Paseo Manuel de Lardiz\'abal 4, 20018 Donostia-San Sebasti\'an, Spain}

\date{\today}

\begin{abstract} 
The recent development of all-electrical electron spin resonance (ESR)
in a scanning tunneling microscope (STM) setup has opened the door
to vast applications. Despite the fast growing number of experimental
works on STM-ESR, the fundamental principles remains unclear. 
By using a cotunneling picture, 
 we show that the spin resonance signal can be explained as a time-dependent
variation of the tunnel barrier induced by the alternating 
electric driving
field.
We demonstrate how this variation translates into the resonant frequency
response of the direct current.  
Our cotunneling theory explains the main experimental findings.
Namely, the linear dependence of the Rabi flop rate with the alternating bias amplitude, the absence of
resonant response for spin-unpolarized currents,
and the weak dependence on the actual atomic species.
\end{abstract}

 \maketitle

\section{Introduction}

The demonstration of reproducible
single-atom~\cite{Baumann_Paul_science_2015} and
single-molecule~\cite{Mullegge_Tebi_prl_2014,Mullegge_Rauls_prb_2015} electron
spin resonance (ESR) has opened new avenues in the analysis of surface
science at the atomic scale. Conserving the atomic spatial resolution of
the scanning tunneling microscopy (STM), STM-ESR provides   unprecedented
energy resolution, in the neV energy scale.~\cite{Willke_Paul_sciadv_2018}
Moreover, it can be combined with high time resolution pump-and-probe
techniques.~\cite{Loth_Etzkorn_science_2010,Paul_Yang_natphys_2017}
This has allowed access to the dipolar interaction between close
magnetic adatoms, GPS-like localization of magnetic impurities
on a surface,~\cite{Choi_Paul_natnano_2017}  single-atom
magnetic resonance imaging,~\cite{Willke_Yang_arXiv_2018} and
spectroscopy,~\cite{Natterer_Yang_nature_2017} probing an adatom
quantum coherence,~\cite{Willke_Paul_sciadv_2018} tailoring the
spin interactions between $S=1/2$ spins,~\cite{Yang_Bae_prl_2018}
measuring and manipulating the hyperfine interaction of
individual atoms~\cite{Willke_Bae_science_2018} and
molecules~\cite{Mullegge_Tebi_prl_2014,Mullegge_Rauls_prb_2015}
 or  controlling the nuclear polarization of individual
atoms.~\cite{Yang_Willke_natnano_2018}

Despite the success of this new experimental technique, there
 are still many open questions about the mechanism leading to
 the all-electric ESR signal. The most prominent question is,
  how can a magnetic moment respond resonantly
 to an AC electric field. Several theoretical proposals have been
 formulated.~\cite{Baumann_Paul_science_2015,Berggren_Fransson_2016,Lado_Ferron_prb_2017,Shakirov_Rubtsov_arXiv_2018}
 Baumann {\it et al.} conjecture~\cite{Baumann_Paul_science_2015}
 that the AC  electric field induces an
adiabatic mechanical oscillation of the adatom, leading to a modulation of the
 crystal field which, together with the spin-orbit, originates spin
 transitions under very particular symmetry constrains.
 A different mechanism could be the phonon excitations induced by the electric field, which efficiently
couples to the magnetic moments as described by Chudnovsky and collaborators.~\cite{Chudnovsky_Garanin_prb_2005,calero_rabi_2007}   
This model has
been successfully applied to explaining the ESR signal in molecular
magnets~\cite{Mullegge_Rauls_prb_2015}.
Unfortunately,
the excitation of unperturbed phonons in MgO/Ag(100) by a driving AC electric
field leads to zero spin-phonon coupling.~\cite{Preliminar}

Berggren {\it et al.}~\cite{Berggren_Fransson_2016} proposed that
 the spin polarization of the electrodes generates a finite time
 dependence of the uniaxial and transverse anisotropy with the
 AC signal. They showed that this change leads to a finite ESR signal
 in integer spins systems, and they predicted a dependence of the
 ESR frequency on the tip-sample distance, a shift that has not been
 observed in recent experiments~\cite{Willke_Paul_sciadv_2018} when changing the current by a factor
 30.
  Lado {\it et al.}~\cite{Lado_Ferron_prb_2017} suggested a
  combination of the distance-dependent exchange with the magnetic
  tip and the adiabatically driven mechanical oscillation of the
  surface spins. However, the amplitude of these oscillations and
  the derived driving strength were too small to account for the
  observation. In addition, this current-related mechanism also seems to be
  in contradiction with the observation of a current-independent Rabi
  flop-rate.~\cite{Willke_Paul_sciadv_2018}
An alternative scenario that does not rely on the coupling to the orbital
(and symmetry dependent) degrees of freedom was introduced by  Shakirov
{\it et al.},~\cite{Shakirov_Rubtsov_arXiv_2018} who defended that the
ESR signal appears as a consequence of the non-linearity of the coupling
between the magnetic moment and the spin-polarized current, which should
yield a strong current dependence, again,  contrary to the 
experimental observations.~\cite{Willke_Paul_sciadv_2018}

Making things more puzzling, not only does a detailed study of the
ESR signal demonstrate a current-independent Rabi induced flop
rate,~\cite{Willke_Paul_sciadv_2018} 
but the ESR signal is observed with virtually all the atomic species
employed with Rabi flop-rates surprisingly
constant: Fe, Ti, Mn, Cu, and Co.~\cite{\expESR}\footnote{Although in the
first work on ESR,~\cite{Baumann_Paul_science_2015} there was not resonant
signal on Co, authors have confirmed us that it can also be detected.}

Using a cotunneling picture of the tunneling current, 
here we show that a frequency dependent DC
current can appear as a modulation of the tunnel barrier in the
STM setup, in the spirit of  the Bardeen theory for the tunneling
current.~\cite{Bardeen_prl_1961} The resulting spin-electron coupling is similar
to the mechanism behind the excitation of molecular
vibrations in the inelastic electron tunneling spectroscopy
conducted with STM.~\cite{Lorente2000,Lorente2004} 
As explained in Ref.~[\onlinecite{Baumann_Paul_science_2015}],
the ESR signal is proportional to the square of
the Rabi flop-rate, and thus, a non-zero
Rabi flop-rate is a necessary condition
to find ESR-active systems. Besides, since the detection
mechanism is based on a magnetoresistive effect,~\cite{\expESR} a strong
Rabi flop-rate is not a sufficient  condition to observe STM-ESR,
and maximum ESR contrast is achieved for a half-metal electrode. By
taking the example of the Fe adatom on MgO, we demonstrate that
the magnitude of this effect is in quantitative agreement with the
experiments~\cite{Baumann_Paul_science_2015,Willke_Paul_sciadv_2018}. In
addition, the proposed
cotunneling picture reproduces the observed voltage 
and current dependences.
We further demonstrate that the mechanism can be applied to explain
the ESR signal of different spins.

The paper is organized as follows. We initially expound
the theoretical model based on time-dependent cotunneling. Next, we
show the results of the theory applied to a simple single-orbital system,
which allows us to explore the physics of the exciting process and the
main ingredients needed to obtain an ESR-active system. We study a
realistic system by computing the ESR signal of a single Fe adsorbate
on a layer of MgO grown on Ag (100), reproducing the main
experimental findings of Refs.~[\onlinecite{Baumann_Paul_science_2015}] and [\onlinecite{Willke_Paul_sciadv_2018}].
In the Discussion section we analyze the main ingredients of
the theory and their implication on the physics of the ESR excitation and
finally, the Conclusions summarize the main findings of this work.

\section{ESR theoretical modeling \label{generalD}}

We model the STM-ESR experimental setup with a time dependent electronic
Hamiltonian,  ${\cal H}(t)= {\cal H}_{\rm r}+{\cal H}_{\rm C}(t) +{\cal
H}_{\rm tun}(t)$, where ${\cal H}_{\rm r}$ correspond to the Hamiltonian
of the reservoirs, considered as free electron gasses, ${\cal H}_{\rm
C}(t)$ models the magnetic adatom and ${\cal H}_{\rm tun}(t)$ corresponds
to the tunneling Hamiltonian that adds or removes an electron from the
magnetic adatom. Notice that we have assumed a time-independent reservoir
Hamiltonian, which ensures a constant occupation of each electrode's single-particle states
 $|\alpha\rangle$,~\cite{Jauho_Wingreen_prb_1994}  where the single particle quantum number $\alpha\equiv (\vec
k,\sigma,\eta)$ labels the electrons in the $\eta$ electrode with $\vec
k$ wavevector and spin $\sigma$. 
  In other words,
the difference in  chemical potentials  $\mu_L-\mu_R=eV$ appearing in
the distribution functions $f(\epsilon)$ is time independent.\footnote{If
the occupations are assumed to change with time, the total number of
electrons in the contact is no longer conserved, leading to a charge
pileup in the contacts. In addition, it also originates an instantaneous
loss of phase coherence in the contacts~\cite{Jauho_Wingreen_prb_1994}.}

In the STM-ESR experiment,~\cite{\expESR} the magnetic atoms are
deposited on a few MgO layers (from 1 to 4 atomic monolayers) on top
of an Ag(100) substrate.  Bulk MgO constitutes a very good insulator
with an energy bandgap of 7.2 eV.~\cite{Taurian_SSCom_1985}
Hence, the coupling between the itinerant electrons on both,
the Ag substrate and the tip can be treated within perturbation
theory.~\cite{Ternes_njp_2015,Delgado_Rossier_pss_2017}
%
The dissipative dynamics of quantum systems weakly coupled
to the environment in the absence of a driving field is well
described by the perturbative Bloch-Redfield (BRF) master
equation.~\cite{Breuer_Petruccione_book_2002} A non-formal
approximate evolution of the reduced density matrix $\hat
\rho(t)$ describing the quantum system in the presence
of an AC driving field can be given in the form of a Bloch
equation.~\cite{Cohen_Grynberg_book_1998,Delgado_Rossier_pss_2017}
Thus, $\hat\rho(t)$ satisfies the following Liouville's
equation,
\beqa
\frac{d\hat\rho(t)}{dt}=-\frac{i}{\hbar}\left[{\cal H}_C(t),\hat\rho(t)\right]+{\cal L}\hat\rho(t),
\label{MMEQ}
\eeqa
where ${\cal L}$ will take the form of a linear Lindblad
 super-operator.~\cite{Breuer_Petruccione_book_2002} ${\cal
 L}\hat\rho(t)$ is responsible for dissipation and thus,
 decoherence and relaxation. In our approach, it will be given
 by the Bloch-Redfield tensor in the absence of the driving
 field.~\cite{Cohen_Grynberg_book_1998,Delgado_Rossier_pss_2017} This
 method will be adequate to describe weak fast-oscillating driving
 fields.~\cite{Hofer_LLobet_njp_2017,Gonzalez_Correa_osid_2017}
 We remark that Eq. (\ref{MMEQ}) leads to the Bloch equations for a
 driven two level system (TLS) where the effective Hamiltonian takes the form
 \begin{equation}
{\cal H}_C^{\rm TLS}(t)=
\begin{pmatrix}
\epsilon_a & \hbar \Omega\cos (\omega t) \\
\hbar \Omega\cos (\omega t) & \epsilon_b
\end{pmatrix}.
\label{HC}
\end{equation}
The diagonal terms are the energy levels of the two
states, $|a\rangle$ and $|b\rangle$, and the off-diagonal term
is the coupling between them. The coupling in a static two-level
system is given by the Rabi flop-rate $\Omega$, see for example
Ref.~[\onlinecite{Cohen_Grynberg_book_1998}].  In the present case, the
AC driving field leads to a modulation of the coupling with the same
frequency as the external field, $\omega/2 \pi$.
A more accurate treatment of the driving term can be obtained using the
 Floquet theory, as implemented for instance in the photon-assisted
 tunneling.~\cite{Stano_Klinovaja_prb_2015}

Equation (\ref{MMEQ})  assumes that the interaction ${\cal
H}_{\rm tun}$ with the reservoirs, included in ${\cal
 L}\hat\rho(t)$, only induces fluctuations around a
zero-average,~\cite{Cohen_Grynberg_book_1998,Breuer_Petruccione_book_2002}
i.e.,  ${\rm tr}_R[\hat \rho_R{\cal H}_{\rm tun}]=0$, where $\hat\rho_R$
is the thermal equilibrium density matrix of the reservoirs and the
trace is over the reservoirs degrees of freedom. Hence, without 
changing the total Hamiltonian, we can add and substract the same quantity, ${\rm Tr}_R[\hat \rho_R{\cal
H}_{\rm tun}]$, and 
 we redefine  the tunneling Hamiltonian as
${\cal H}_{\rm tun}'={\cal H}_{\rm tun}-{\rm Tr}_R[\hat \rho_R{\cal
H}_{\rm tun}]$ and the system Hamiltonian
\beq
{\cal H}_C'={\cal H}_C+{\rm Tr}_R[\hat \rho_R {\cal H}_{\rm tun}].
\label{Hrenormalized}
\eeq
For notation clarity, we omit the primes and, unless otherwise stated, we will refer to the renormalized Hamiltonians.  

\textit{Multilevel adatom.-} To explore the origin
of the ESR signal in a realistic multilevel system, we concentrate on the Rabi
frequency, $\Omega$. In particular, we focus on the situation where
the driving frequency $\omega$ is close to the Bohr frequency of the
transition between the first excited state, $|b\rangle$, and the ground
state, $|a\rangle$, $\omega_{ba}=(E_b-E_a)/\hbar$,
while all other transitions are far away.
We follow the same procedure leading to the Bloch equations, but now we
consider that the system Hamiltonian ${\cal H}_C$ has an arbitrary number
of states. 
Hence, we define  static and driving
parts, ${\cal H}_C(t)={\cal H}_C^0+\delta{\cal H}_C\cos(\omega t)$
following the two-level scheme, Eq.~(\ref{HC}).
Using a similar notation, we write the time-dependent interaction 
as ${\cal H}_{\rm tun}(t)\equiv {\cal H}_{\rm tun}^0+\delta {\cal H}_{\rm tun} \cos(\omega t).$

The transition rates $\Gamma_{nn'}$ between any two states
$n$ and $n'$ of  ${\cal H}_C^0$ can be calculated with the
standard expressions, reproducing the Fermi's Golden rule
results,~\cite{Breuer_Petruccione_book_2002,Cohen_Grynberg_book_1998}
and similarly for the decoherence rates $\gamma_{nm}$ between any
 two pair of states $n$ and $m$.~\cite{Delgado_Rossier_pss_2017}
The Rabi flop-rate is defined by the off-diagonal matrix elements of ${\cal H}_{\rm C}(t)$.
Then, following Eqs. (\ref{HC}) and (\ref{Hrenormalized}), we can write
\beq
\hbar\Omega \approx 
\langle a|\delta{\cal H}_{\rm C}|b\rangle+\langle a|{\rm Tr}_R\left[\hat \rho_R\delta {\cal H}_{\rm tun}\right] |b\rangle.
\label{DefOmega}
\eeq
Coupling of a quantum system with a reservoir also induces a
renormalization of the system's energy levels proportional to the square
of the interaction.~\cite{Breuer_Petruccione_book_2002,Anderson_prl_1966} 
This also yields a time-dependent contributions quadratic in the tunneling term and linear in $\delta {\cal H}_{\rm tun}$, being effectively a third order correction
to the decoupled system. Thus, we will neglect it for consistency.

\subsection{Description of the STM junction as a tunnel barrier \label{Bardeen}}
The driving electric field can in principle translate into two effects. First, a modulation of the tunneling amplitudes
$V_{\alpha,{\bf i}}(t)$  describing the (spin-conserving) hopping
between the adatom state, given by ${\bf i}\equiv (\ell,\sigma)$ (with $\ell$ the
orbital and $\sigma$ the spin degrees of freedom of
the atomic levels) and the reservoir
state $|\alpha \rangle$. Second, a time dependence of the adatom's energy levels, similar to a Stark
energy shift of the d-shell. 
 For a single level Anderson model, by using
a Schrieffer and Wolff transformation,~\cite{Schrieffer_Wolff_pr_1966}
one can demonstrate that both time-dependences can be treated as a time-dependent exchange coupling between
the adatom spin and the reservoirs spin density, in the spirit of
Ref.~[\onlinecite{Lado_Ferron_prb_2017}]. 
Density functional theory calculations
for the Fe/MgO/Ag(100) system show that the adatom's level shifts are negligible under
an external electric field.~\cite{DFT}
This implies the prevalence of the modulation of
the tunnel barrier, similar to the case of inelastic tunneling spectroscopy 
(IETS).~\cite{Lorente2000,Lorente2004}
Then, we assume that the adatom Hamiltonian is not affected
by the AC driving field, i.e., $\delta{\cal H}_{C}=0$.

We model the effect of the AC driving field on the tunneling amplitudes
as follows.  We assume that the STM junction can be treated as a square
vacuum barrier of length $L$ and height $U$, and that the tip and sample
have the same work functions.~\footnote{Tip is usually
made of W, but indentation may lead to different apex atoms, while
metallic substrate is Ag(100).}  The AC applied voltage leads to a
time-dependent change of the transmission amplitude. For small-enough bias, we approximate
\begin{equation}
 V_{\alpha,{\bf i}}(t) \approx V_{\alpha,{\bf i}}^0\left(1+eV_{ac}\cos(\omega
t)|\delta_{ac}|^{-1}\right).  
\label{hybrit}
\end{equation}
Following a WKB description, and
introducing the wavenumbers $k=\sqrt{2m^* \epsilon}/\hbar$
and $\kappa=\sqrt{2m^* (U-\epsilon)}/\hbar$ we have
that~\cite{Messiah_book_1999} 
\beq \delta_{ac}^{-1}\approx
\frac{m^*\left(k_F(1-L\kappa_F)-iL\kappa_F^2\right)}{\hbar^2
\kappa_F^2(k_F+i\kappa_F)}.  
\eeq 
Here we have assumed that $k$ and
$\kappa$ can be approximated by their values at the Fermi level. This
is adequate to describe the tunneling current under the experimental low
bias conditions.~\cite{Delgado_Rossier_pss_2017}

The Rabi flop-rate can be evaluated using Eqs. (\ref{DefOmega}) and (\ref{hybrit}). Then we have that
\beq
\hbar\Omega\approx \left|\frac{eV_{ac}}{\delta_{ab}}
\langle a|{\rm Tr}_R\left[\hat\rho_R H_{\rm cot.}^0\right]|b\rangle \right|.
\label{OmeB}
\eeq

\subsection{Cotunneling transition amplitudes}
We now use a description based on second order cotunneling
transport~\cite{Wegewijs_Nazarov_arXiv_2001,Delgado_Rossier_prb_2011}
adapted to the time-dependent Hamiltonian ${\cal H}(t)$. The central idea
is that, as the adatom can be considered within the Coulomb blockade
regime, where charging is energetically costly, we
restrict the atomic configurations to the ones with $N_0$ and $N_0\pm 1$ electrons.
This will allow us to substitute the tunneling Hamiltonian ${\cal H}_{\rm
tun}(t)$, where the adatom  charge fluctuates, by an effective cotunneling
Hamiltonian $H_{\rm cot.}(t)$ acting only on the $N_0$ charge-space.
The approximation will be valid as long as the system is far from
resonance, i.e., $|E_{\pm}-E_0\pm E_F|\gg k_BT, |eV|$, where $E_\pm$
($E_0$) are the ground state energies of the system with $N_0\pm 1$ ($N_0$)
electrons, while $V$ is the applied bias voltage and $T$ the temperature.

The effective cotunneling Hamiltonian in the absence of driving field
can be found in Ref.~[\onlinecite{Delgado_Rossier_prb_2011}]. The
details of the derivation for a time-dependent tunneling are given in
Appendix~\ref{AppendixC}. Thus, one can write it as
%
\beq
H_{cot.}(t)\approx \sum_{\alpha\alpha'}
\left[\hat{T}_+(\alpha\alpha';t)f_\alpha^\dag f_{\alpha'}+\hat{T}_-(\alpha\alpha';t)f_\alpha f_{\alpha'}^\dag\right]
\label{Hcotun_eff}
\eeq
where $\hat{T}_\pm(\alpha\alpha';t)$, given by
Eqs. (\ref{Tminus}-\ref{Tplus}), denote (time-dependent) transition
amplitude matrix elements.

The evaluation of the Rabi frequency, transition rates and decoherence
rates, requires to calculate the transition amplitudes $\hat
T_\pm(\alpha,\alpha',t)$ introduced in Eq. (\ref{Hcotun_eff}).

In the following, we denote by $|m^\pm\rangle$ and $E_{m^\pm}$ the
eigenvectors and eigenvalues of the adatom Hamiltonian ${\cal H}_C$
with $N_0\pm 1 $ electrons, while  $|m\rangle$ and $E_{m}$ will be
used for the the eigenvectors and eigenvalues of the $N_0$-electron configuration.  Thus, using Eq. (\ref{DefOmega}) and Eqs. (\ref{Tminus}-\ref{Tplus}),
we have that
\begin{widetext}
\beq
\Omega =  
\frac{ -e V_{ac}}{\hbar |\delta_{ab}|}
\sum_{\alpha}
\left[\sum_{m_-}\frac{f^-_{\eta_\alpha}(\epsilon_\alpha)\Lambda_{m^-,\alpha}}{\left(\Delta E_{m_-} -\mu_-+\epsilon_\alpha\right)}
 +\sum_{m_+}\frac{f^+_\eta(\epsilon)\Lambda_{m^+,\alpha}}{\left(\Delta E_{m_+} +\mu_+ -\epsilon_\alpha\right)}
 \right],
\label{OmegaG3p}
\eeq
\end{widetext}
where we have introduced the excitation 
energies $\Delta
E_{m^\pm}=E_{m^\pm}-E_{0^\pm}$ and the charging energies of the adatom 
$\mu_{+} = E_{0^+}-E_0$ and $\mu_{-} =
E_{0}-E_{0^-}$. Here $f^+_\eta(\epsilon)=f(\epsilon-\mu_\eta)$ and
$f^-_\eta(\epsilon)=1-f(\epsilon-\mu_\eta)$, with $f(\epsilon)$ the
Fermi-Dirac distribution and $\mu_\eta$ the chemical potential of the
$\eta$-electrode. In addition, we have defined
\beq
 \Lambda_{m^\pm,\alpha}=\sum_{\ell\ell'}V_{\alpha,\ell'} V_{\alpha,\ell}^*\gamma_{ab}^{m^\pm}(\alpha\ell',\alpha\ell),
 \label{d_Lambda}
\eeq
where $\gamma_{ab}^{m^-}(\alpha\ell',\alpha\ell)=\langle
a|d_{\ell\sigma}^\dag |m_-\rangle \langle m_-|d_{\ell'\sigma}|b\rangle$
and $\gamma_{ab}^{m^+}(\alpha\ell',\alpha\ell)=\langle a|d_{\ell\sigma}
|m_+\rangle \langle m_+|d_{\ell'\sigma}^\dag|b\rangle$.
Equation (\ref{OmegaG3p}) is the central result of this work. 
Notice that contrary to what happens in the calculation of transition and decoherence
rates,~\cite{Delgado_Rossier_prb_2011,Delgado_Rossier_pss_2017} here the
energies $\epsilon_{\alpha}$ are not limited to a small energy window
around the Fermi level. Thus, the evaluation of Eq. (\ref{OmegaG3p})
requires a precise knowledge of the hybridization functions
$V_{\alpha,\ell'}$.

For convenience, we introduce the density of states, 
$\rho_\eta(\epsilon)=\sum_{k\sigma}\delta(\epsilon-\epsilon_{k\sigma})$
and the spin polarization of the electrode, ${\cal
P}_\eta(\epsilon)=(\rho_{\eta}^{\rm Max}(\epsilon)-\rho_{\eta}^{\rm Min}(\epsilon))/\rho_\eta(\epsilon)$, with 
$ \rho_{\eta}^{\rm Max}$ ($ \rho_{\eta}^{\rm Min}$) the majority-spin (minority-spin) density of states.

\subsection{Current detection of the STM-ESR } 

In all the experiments realizing STM-ESR,~\cite{\expESR} the detection
frequency bandwidth is around 1 kHz, so the driving AC voltage,
modulated on the GHz frequency range, is averaged out. Thus, the
resonant signal is detected only by the magnetoresistive static
current. The resulting DC current can be evaluated in terms of the transition
rates $\Gamma_{mm'}^{\eta\eta'}$ ($\eta\ne \eta'$) between the $\eta$ and
$\eta'$ electrodes and the non-equilibrium occupations $P_{m}(V,\omega)$:
 \beqa
 I(V,\omega)=e\sum_{mm'}P_{m}(V,\omega)\left(\Gamma_{m,m'}^{T,S}-\Gamma_{m,m'}^{S,T}\right).
 \label{current}
 \eeqa
Here, the non-equilibrium occupations $P_{m}(V,\omega)$ will be the
result of a stationary condition $d\hat\rho(t)/dt=0$ that
defines the steady state, and it accounts for the coherence between
the $|a\rangle$ and $|b\rangle$ states connected by the ESR signal.\footnote{In the
Bloch-Redfield theory, there are other terms of the Redfield tensor
proportional to the coherences $\rho_{mm'}$ that may also contribute to
the current.~\cite{Breuer_Petruccione_book_2002} However, they involve
rates coupling coherences with occupations, which are usually quite
small and will be discarded here.}

\begin{figure}
\centering
\includegraphics[width=1.\linewidth,angle=0]{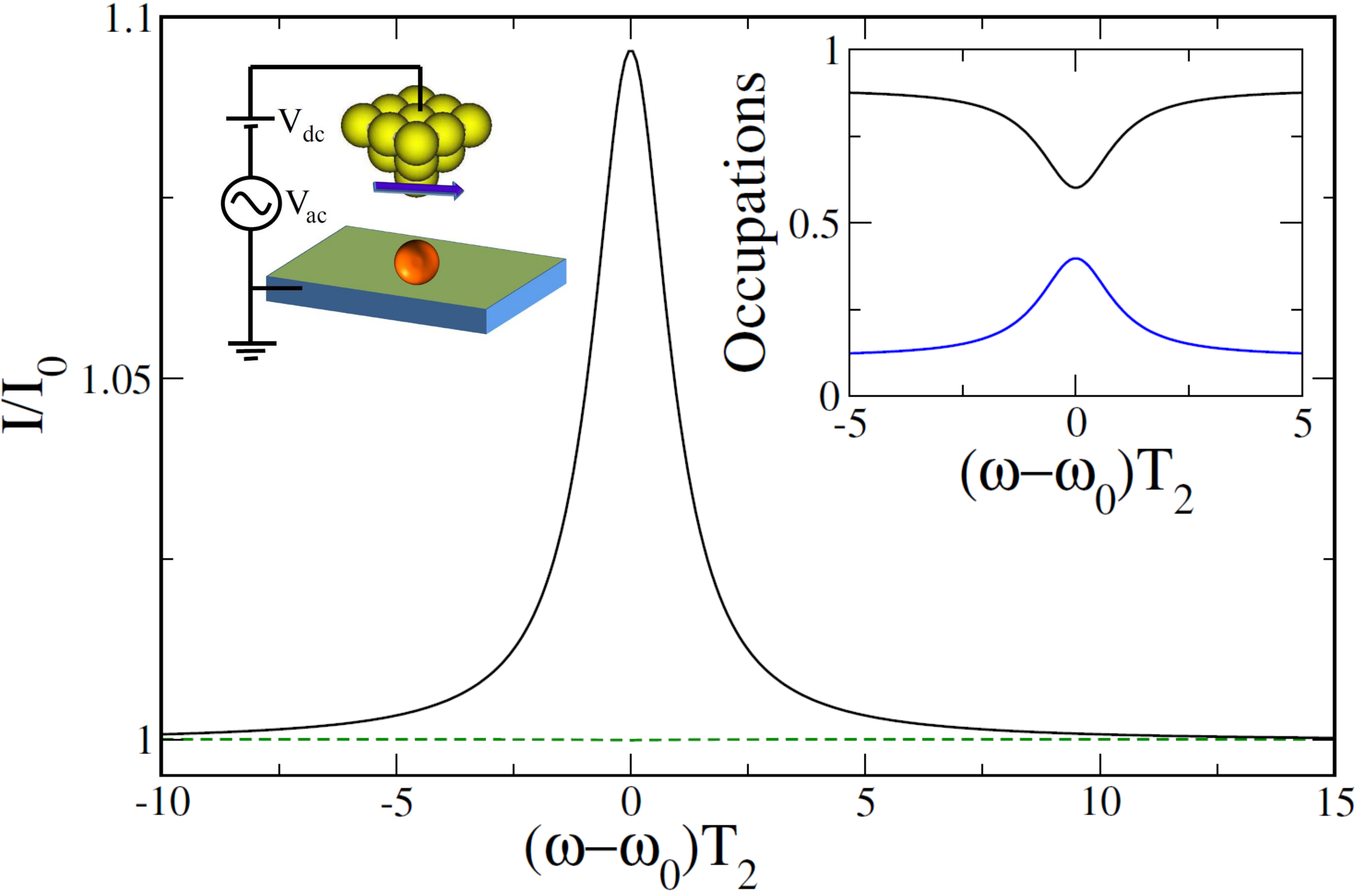}
\caption{ 
Current detection of STM-ESR. Variation of the DC current as a
function of the detuning $\omega-\omega_0$ in units of $1/T_2$, with
$\omega_0=(E_b-E_a)/\hbar$ for ${\cal P}_T=0.33$ (black) and  ${\cal
P}_T=0$ (green dashed line). The current is given in terms $I_0$, the DC current far from resonance. (Left inset) Scheme of the STM-ESR setup: a radiofrequency bias voltage is applied in addition to the DC voltage between the spin-polarized tip and the surface. 
(Right inset) Variation of the occupation of the ground state, $|a\rangle$, (black) and first
excited state, $|b\rangle$, (blue) with detuning. The different parameters
are chosen to match the conditions of Fig.3(C) in Baumann {\it et
al.}~\cite{Baumann_Paul_science_2015} with $V_{ac}=8$ mV.
}
\label{fig0}
\end{figure}

Equation (\ref{current}) makes explicit the working mechanism of the
STM-ESR: the occupations $P_{m}(V,\omega)$ respond to the driving
frequency and the changes are reflected in the DC current
$I(V,\omega)$. The consequences on the current and occupations can be
seen in Fig.~\ref{fig0}. This figure illustrates the magnetoresistive
detection mechanism. Here we have used a two-level description where
the steady-state density matrix is given by the  analytical solutions
of the Bloch equations,~\cite{Delgado_Rossier_pss_2017} 
which is determined by the relaxation time $T_1=1/\Gamma_{ab}$, the decoherence time $T_2=1/\gamma_{ab}$ and the Rabi flop rate $\Omega$.
 We have used the parameters extracted from Baumann {\it
et al.}:~\cite{Baumann_Paul_science_2015} $T_1\approx 88\; \mu$s,
$T_2\approx 200$ ns, $\Omega\approx 2.6$ rad/$\mu$s. In addition, 
we take a
set-point current of $I= 0.56 $ pA at $V= 5 $ mV,  while we assume a tip polarization
 ${\cal P}_T=+0.33$.
The finite tip polarization leads to a magnetoresistive response: the electrons' tunneling rates depend on the relative orientation between the local spin and the tip magnetization, together with the sign of the applied bias.\cite{Delgado_Palacios_prl_2010,Loth_Bergmann_natphys_2010} 
Furthermore,  close
to the resonant frequency, the occupations of the two low-energy states
tends to equilibrate,  as observed in the inset.
 This change of $P_{m}(V,\omega)$ is then reflected
as a change in the DC current detected by the STM.

\section{Results\label{Results}} 

In Sec.~\ref{generalD} we have sketched the cotunneling mechanism
leading to the STM-ESR. In our description, the consequence of the
ac driving voltage is summarized in the non-equilibrium occupations
$P_{m}(V,\omega)$ and, more explicitly, on the Rabi flop rate $\Omega$,
given by Eq.(\ref{OmegaG3p}).  In order to illustrate the results, we
make a quite strong simplification: we assume an energy-independent
hybridization $V_{\alpha\ell}\equiv V_{\eta,\ell}$ and density of
states $\rho_{\eta}(\epsilon)\approx \rho_\eta$ and, consequently, we introduce an energy cut-off $E_c$. This raw
approximation will enable us to estimate the Rabi flop-rate and the ESR
current response. By comparing the results with the experimental ones,
we show that despite the  approximations, the predicted behavior is in
qualitative agreement. On the down side, our approach overestimate
the Rabi frequency by one order of magnitude.

Below we work out the explicit expressions of the Rabi frequency and we
illustrate the main results for two cases, a single orbital Anderson
Hamiltonian and the multiorbital case describing the Fe/MgO/Ag(100)
system.  In the former, the only ingredients are he charging energy of the adatom and the induced Zeeman splitting. In the second case, we describe the magnetic adatom by a multiorbital
Hubbard model that includes the Coulomb repulsion between the
impurity d-electrons, the crystal field calculated by a point-charge
model,~\cite{Dagotto_book_2003} the spin-orbit coupling and the
Zeeman term.~\cite{Delgado_Rossier_prb_2011,Ferron_Delgado_njp_2014}
In doing so, we assume hydrogenic-like wavefunctions for the Fe orbitals. The Coulomb interaction
is parametrized by a single parameter, the average on-site repulsion $U$.
The resulting crystal field depends on two parameters,
the expectation values  $\langle r^2\rangle$ and $\langle
r^4\rangle$,~\cite{Dagotto_book_2003} while the spin-orbit coupling
will be defined by its strength $\xi_{SO}$. Despite the quantitative
limitations of the point-charge models, they provide a good description
of the symmetry of the system and they are very often used to describe
ESR spectra.~\citep{Abragam_Bleaney_book_1970}

\subsection{Single-orbital Anderson model \label{soAm}}
We start discussing the simplest model for a magnetic impurity: the single-orbital Anderson model. This model, which was introduced to describe magnetic
impurities on a non-magnetic metal host,~\cite{Anderson_prl_1966}
is equivalent to a single $S=1/2$ spin exchange coupled to conduction electrons.~\cite{Schrieffer_Wolff_pr_1966}
Then, it may be used as an idealization
of the STM-ESR experiments on hydrogenated Ti atoms on 
MgO.~\cite{Yang_Bae_prl_2018,Willke_Bae_science_2018} 
The $S=1/2$ spin is isotropic, and the matrix elements
$\gamma_{ab}^{m^\pm}(\sigma\sigma)$ can be evaluated analytically.

Let us consider that the system is under the influence of a static
 magnetic field $B_x$, so that $|a\rangle$ and $|b\rangle$  are
 eigenvectors of the spin operator $\hat S_x$  and $g\mu_B B_x =\hbar
 \omega_0$ is the Zeeman splitting. Then, assuming that only the tip is spin polarized and using
the same notation as in Eq.~(\ref{OmegaG3p}),  one gets after some straightforward algebra that
 $\sum_{\sigma}(1+\sigma {\cal P}_\eta)\Lambda_{m^\pm,\eta\sigma}\equiv
 \mp V_T^2 {\cal P}_T\delta_{\eta,T}$, where $V_T$ is the hopping between the single level and the tip. In other  words, only coupling with a spin-polarized electrode gives a finite contribution to $\Omega$.\footnote{Notice that the spin-dependent excitation energy
 in the denominators of Eq. (\ref{OmegaG3p}) can be neglected when
 the addition energy is very large.} 
When the extension of the hybridization function $\Gamma_T (E)=
2\pi  \rho_T (E) V_T^2$, given by the cuttof $E_c$,
is much larger than the thermal energy $1/\beta$, 
one gets
\beq
\Omega  \approx   
\frac{e\Gamma_{T}\left| V_{ac}{\cal P}_T\right|}{ h|\delta_{ab}|}
\left| {\cal I}^-(\mu_-,E_c,eV)-{\cal I}^+(\mu_+,E_c,eV)
\right|,
\label{OmegaG3cte}
\eeq
where we have approximated the hybridization function $\Gamma_T$ by its value at the tip Fermi level, $e$ is the elementary charge and the functions ${\cal I}^\pm(\mu_\pm,E_c,eV)$ are defined in 
 Appendix~\ref{AppenI}.


Crucially, the result above relies on the fact that the tip polarization
is normal to the magnetic field producing the Zeeman splitting,
leading to a finite mixing between the eigenvectors $|a\rangle$ and
$|b\rangle$. This should not be surprising since in the standard ESR
protocols,~\cite{Abragam_Bleaney_book_1970} the AC magnetic
field is applied perpendicular to a large static field. In our case,
the AC electric field yields an effective oscillating magnetic field
along the tip polarization direction $z$, which is on resonance with
the Zeeman splitting produced by the applied static magnetic field $B_x$.

The result (\ref{OmegaG3cte}) has a different reading: the proposed 
mechanism does not need any particular anisotropy. 
The key ingredient is thus the effective magnetic field
created by the spin-polarized tip, $B^{\rm eff}=2\hbar\Omega/g\mu_B$,
which is oriented along the tip-polarization direction. 
In order to
have an active ESR signal, this effective field must have a component
perpendicular to the 
the field inducing the Zeeman splitting.

\subsubsection{A single-orbital multispin model}
In general, transition metal adatoms entails
$S \ge 1/2$ spins, and thus, are also subjected to magnetic anisotropy. The dominant interaction with
their surroundings takes the form of an exchange
coupling,\cite{Lorente_Gauyacq_prl_2009,Rossier_prl_2009} which determines the IETS, the
spin relaxation and decoherence.~\cite{Delgado_Rossier_pss_2017} Hence, the total spin, given by the sum of the local spin and scattering electrons spin, is conserved. Thus, we can
model this interaction in the cotunneling context
by considering the scattering of the itinerant electrons with a localized magnetic impurity described by a 
single-orbital state, with a spin $S > 1/2$ [multiplicity $(2S + 1)$] in its $N_0$ electrons state, and spin $S_\pm = S-1/2$.~\footnote{The spin of the $N_0 \pm 1$ electrons state could be either
$S + 1/2$ or $S − 1/2$. However, the conclusions are not affect
by this change.} 
We assume that the states with $N_ 0 \pm  1$ electrons are degenerate, which translates into  $\Delta E_{m_\mp}\mp \mu_\pm= \pm \mu_\pm$. This description has already been used to describe dynamics and IETS of magnetic adatoms adsorbed on thin insulating
layers.~\cite{Lorente_Gauyacq_prl_2009,Gauyacq_Novaes_prb_2010} For simplicity,
we consider that the $(2S + 1)$ states of the system
with $N_0$ electrons will be equally coupled to the tip and surface states.

The model sketched above allows us us to describe the effective exchange interaction $J_{\alpha\alpha'}$ in terms of the transition amplitude operators $\hat T_\pm (\alpha\alpha' )$. In addition, it permits relating the Rabi flop rate, Eq. (\ref{OmegaG3p}), with the
local spin $S$. While the energy dependency is the
same that appears in the single Anderson model, the crucial differences are associated to $\sum_{m_\pm,\sigma}(1+\sigma P_T)\Lambda_{m_\pm,\alpha}$, see Eq. (\ref{d_Lambda}). Using the properties of the Clebsch-Gordan coefficients, one can arrive to~\cite{Sakurai_book_1995}
\beqa
\sum_{m_\pm,\sigma}(1+\sigma P_T)\Lambda_{m_\pm,\alpha}=\frac{P_T}{2S(2S+1)}\langle a|\hat S_z|b\rangle.
\eeqa
Thus, our model predicts  
a linear dependence with the atomic spin, in good agreement with the observation of 
STM-ESR weak dependence on the atomic species.~\cite{\expESR}

An important detail of our results is that the Rabi flop rate is proportional to the tip polarization and the hybridization ${\cal P}_T\Gamma_T$. Hence, it leads to $\Omega\propto I$, which is in apparent contradiction with the experimental observation of a Rabi flop independent of the DC current for the Fe/MgO.~\cite{Willke_Paul_sciadv_2018} With this in mind, we examine below the corresponding results based on a multiorbital Hubbard model.

\subsection{The Fe/MgO/Ag(100) system}
Although STM-ESR has been demonstrated on a variety of
magnetic adatoms,~\cite{\expESR} the most studied system is
Fe/MgO/Ag(100).~\cite{Baumann_Paul_science_2015} In
a recent work, we have demonstrated that this system can
be correctly described by a multiorbital Hubbard model
where the crystal and ligand field was estimated from a DFT
calculation.~\cite{DFT}  Yet, the results were in
qualitative agreement with those obtained with a simpler point-charge
model.~\cite{DFT,baumann2015investigation} Thus here we use
the point charge model results of Ref.~[\onlinecite{DFT}].
\begin{figure}
\centering
\includegraphics[width=1.\linewidth,angle=0]{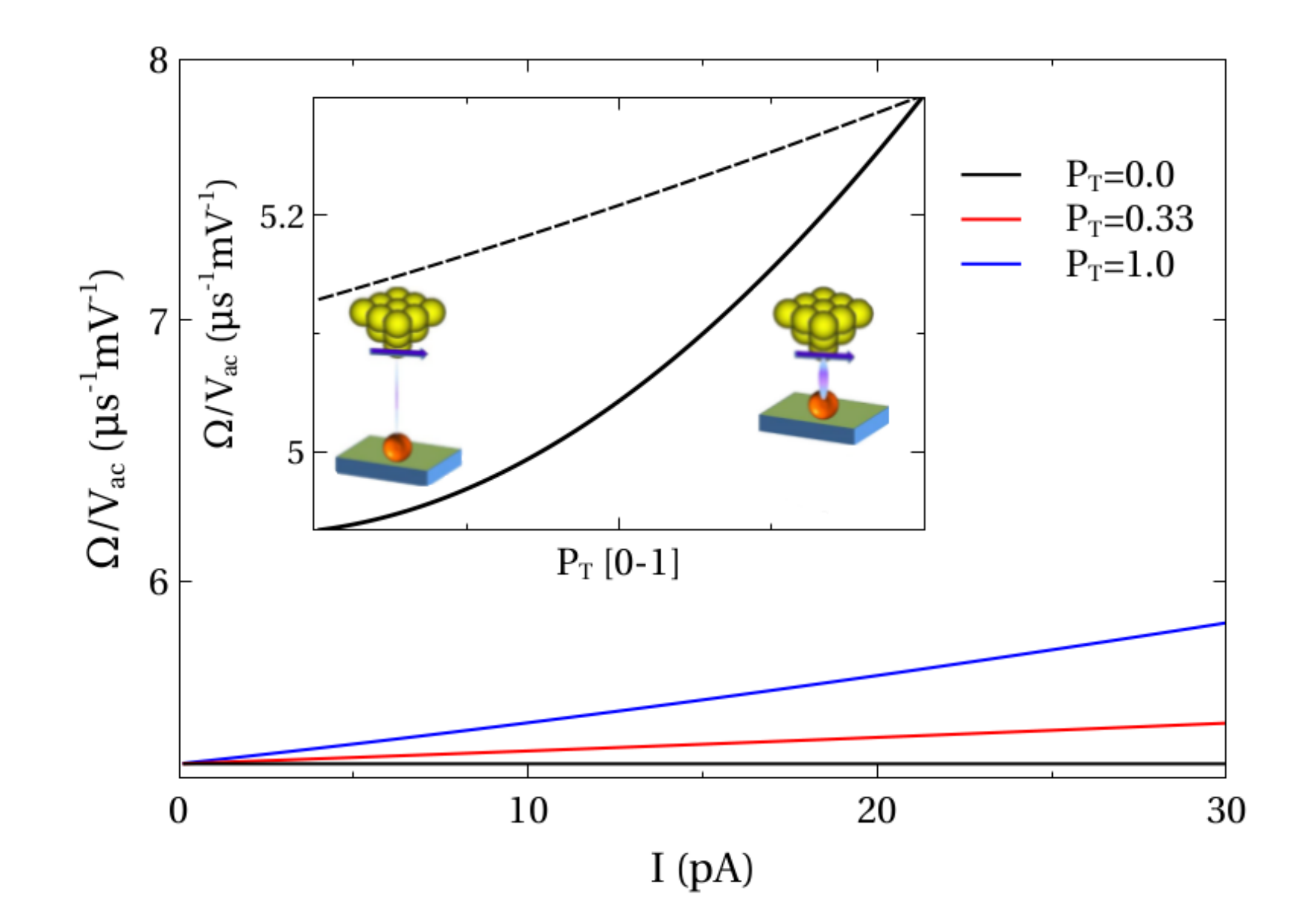}
\caption{ 
Estimation of $\Omega/V_{ac}$ for the Fe/MgO/Ag(100) system
versus the DC current for a constant applied bias voltage $V_{dc}=60$
mV. Each curve corresponds to a different tip
polarization ${\cal P}_T$. The inset shows the dependence with the
tip polarization for the {\it high} (thick solid line) and {\it low}
(thin dashed-line) DC voltages.
}
\label{fig1}
\end{figure}

For the Ag(100) surface we have~\cite{Ashcroft_Mermin_book_1976} that
 $m^*=0.99m_e$ and $k_F\approx \kappa_F\approx 1.1$ \AA$^{-1}$. Typical
 tunneling current measurements are given in a range where $\kappa L\sim
 3-20$, which translates into $|\delta_{ab}^{-1}|\sim (0.3-2.1)\times
 10^{-3}$ meV$^{-1}$. For simplicity, we assume that all Fe-d orbitals
 are equally coupled to the substrate, with an energy broadening
 $\Gamma_S\equiv 2\pi \rho_S |V_s|^2$. In the case of coupling to the
 tip, we assume that only the $d_{z^2}$ is actually coupled, as expected
 from the symmetry of the orbitals, with an induced energy broadening
 $\Gamma_T\propto I$.

In order to check our model, we first take  $\Gamma_S=2.314$ eV to
fit the decoherence time, obtaining $T_2=210 $ ns for the conditions
of Fig. 3C of Ref.~[\onlinecite{Baumann_Paul_science_2015}] at a
driving voltage of $8$ mV. Our cotunneling description then predicts
a relaxation time $T_1$ of the Zeeman-excited state of $T_1=8.72$
ms, to be compared with the experimentally determined $T_1^{\rm
Exp}=88\;\mu$s.  The disagreement between both values can
have two origins. On one side, we have the limitations due to the
oversimplified point-charge model, together with the critical and
different dependences of $T_1$ and $T_2$ on the magnetic anisotropy
parameters. On the other side, this transition may also be mediated by
the spin-phonon coupling.~\cite{Paul_Yang_natphys_2017}  Fortunately,
our STM-ESR mechanism does not strongly depend on $T_1$.

We now turn our attention to the Rabi flop-rate, evaluated according
to Eq. (\ref{OmegaG3p}). The energy integration is done as in
Eq. (\ref{OmegaG3cte}), and the only difference comes from the matrix
elements  $\Lambda_{m^\pm,\eta\sigma}$. In this case, the sums over
$m^\pm$ are extended over all states needed to guarantee the convergence.
Figure \ref{fig1} shows the DC current dependence at
$V_{dc}=60$ mV of the Rabi flop-rate for three different
tip-polarizations: ${\cal P}_T=0,\;0.33$ (close to the one
estimated experimentally~\cite{Willke_Paul_sciadv_2018}) and $1$,
the ideal half-metal case. As observed, especially for intermediate
polarizations, $\Omega$ is barely affected by the current. This
striking result is in agreement with the experimental findings that
shows a current-independent Rabi flop rate for currents between  10
pA and 30 pA,~\cite{Willke_Paul_sciadv_2018} where authors found that
$\Omega/V_{ac}\approx 0.375$ rad.$\mu$s$^{-1}$.

\begin{figure}
\centering
\includegraphics[width=1.\linewidth,angle=0]{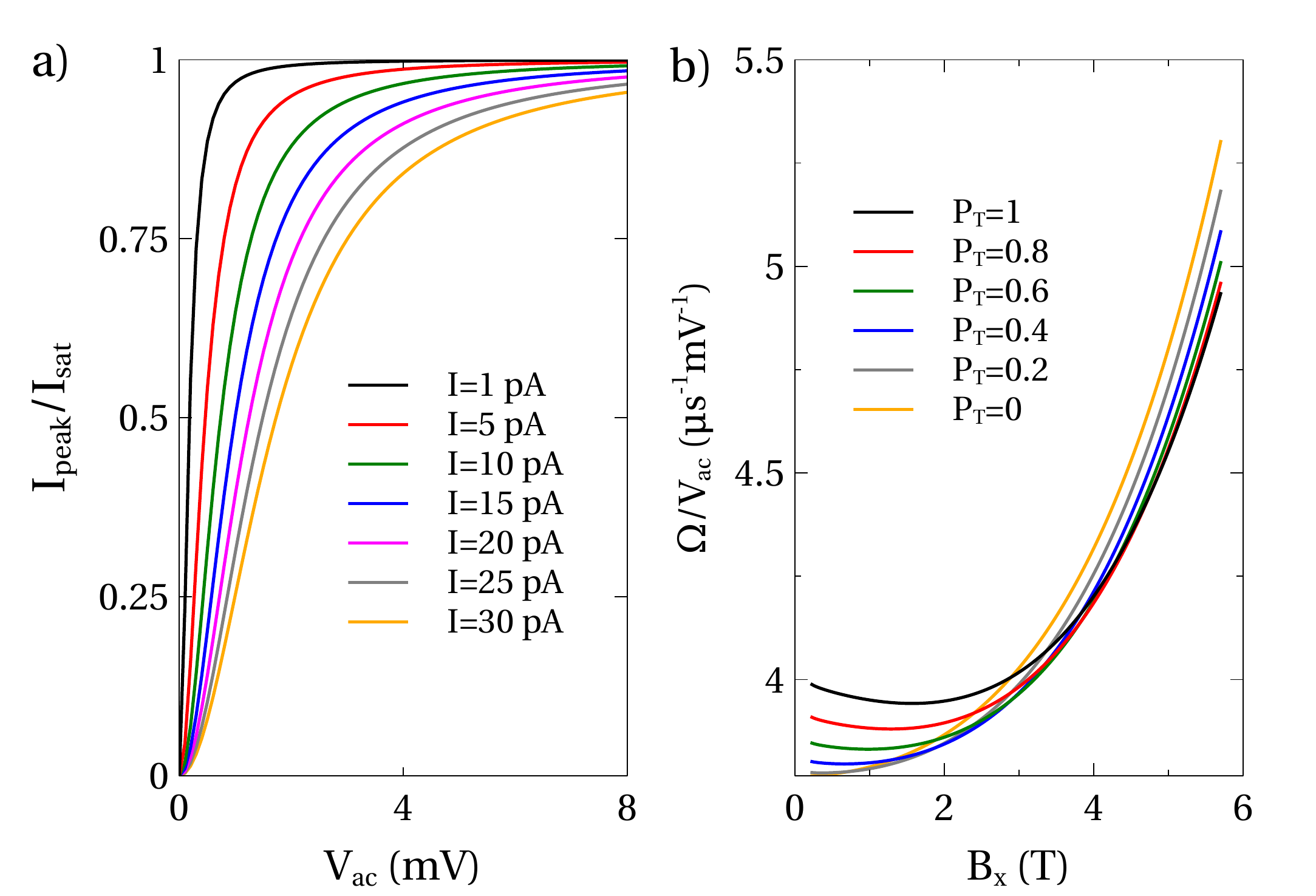}
\caption{ a) Predicted driving function $I_{peak}/I_{sat}$ for the Fe/MgO/Ag(100) system for different currents, to be compared with Ref.~[\onlinecite{Willke_Paul_sciadv_2018}].  
b) Variation of the Rabi flop-rate, $\Omega$, with the external in-plane field $B_x$ for different values of tip polarization.
The field is applied forming an angle $\theta=3.51\times 10^{-2}$ with the surface plane,  with  $B_z=0.2$ T, while $I=30$ pA and $V_{DC}=60$ mV.  
}
\label{fig2}
\end{figure}

The result above points to a crucial ingredient that is not accounted for in the single-orbital Anderson model: the complex orbital structure of the adatom. According to Eq. (\ref{d_Lambda}), electrons tunneling into different orbitals $\ell$ of the adatom will lead to unequal contributions to the Rabi flop-rate. 
The direct consequence is that, contrary to the single-orbital case, the spin averages $\sum_{m_{\pm},\sigma}\Lambda_{m^\pm,\eta\sigma}$ remains finite, which translates into a finite Rabi flop rate at zero current polarization, see inset of Fig.~\ref{fig2}. 
The weak current dependence appears then as a direct consequence: $\Omega$ contains a fix contribution associated to hybridization with the surface, proportional to $\Gamma_S$, and another one of the tip, proportional to $\Gamma_T$ ($\propto I$). Since $\Gamma_S\gg \Gamma_T$ except for very high conductances,\cite{Paul_Yang_natphys_2017} the current independent contribution generally dominates. 
Comparing the polarization dependence
for  {\it low voltage},  with a current set-point of 0.56
pA at $V_{dc}=5$ mV,
 and {\it high voltage}, with a current of 30 pA at $V_{dc}=60$
mV,
we notice that the Rabi frequency is not strongly affected by the dynamics of the excited spin states. 

A key issue is the apparent contradiction of our finite Rabi flop-rate
for zero-polarization with the observation of the STM-ESR signal only
when a spin-polarized tip is used. The solution to this apparent discrepancy is
in the detection mechanism of the ESR: current magnetoresistance. This
is illustrated in Fig.~\ref{fig0}, where we have added the frequency
response when a spin-averaging tip is used, assuming exactly the same
Rabi flop rate. The resulting steady state current is independent of
the frequency and thus, there is not STM-ESR signal.

Willke {\it et al.} analyzed in detail the role of the different
parameters that controls the STM-ESR.~\cite{Willke_Paul_sciadv_2018}
In particular, they observed that the resonant peak current saturates
with the radio frequency voltage $V_{ac}$, both for small and
large set-point currents. In fact, they found that the ratio $I_{\rm
peak}/I_{\rm sat}=\psi(V_{ac})$, which they called {\it drive function},
was given by $\zeta^2/(1+\zeta^2)$ with $\zeta=V_{ac}/V_{1/2}$,
where $V_{1/2}=(T_1T_2)^{-1/2}V_{ac}/\Omega$  is defined as the
half-saturation voltage. The relevance of this drive function is that,
the larger the drive function, the larger the ESR signal, making the
detection more efficient.  Thus, we show in Fig.~\ref{fig2}a) the drive
function obtained from our model, which should be compared with Fig. S2C
of Ref.~[\onlinecite{Willke_Paul_sciadv_2018}].
 Our theory correctly reproduce the general trend with the tunnel
 current. However, due to the overestimation of $T_1$ and $\Omega$,
 our estimated  driving function saturates at lower AC bias voltages.

Finally, we would like to call the attention on one point. The
experimental observation of the STM-ESR signal requires
a finite in-plane magnetic field  $B_x$.~\cite{\expESR} In
Ref.~[\onlinecite{Baumann_Paul_science_2015}], authors argue that this
field introduces a mixing between the states $a$ and $b$, the same
argument that we exploited in our $S=1/2$ spin model of Sec.~\ref{soAm}.
From our expression of the Rabi flop-rate, Eq. (\ref{DefOmega}), we
see that its effect is the same as the
one produced by a transversal AC magnetic field
$B_{\perp}^{\rm eff}=2\hbar\Omega/g\mu_B$ on a $S=1/2$
spin system under the action of an static field $B_{\parallel}^{\rm
eff}=\hbar\omega_{ba}/g\mu_B$.

Hence, the application of a static field that mixes the zero-field
states $|a^0\rangle$ and $|b^0\rangle$, as it is the case of the
Fe/MgO,~\cite{Baumann_Paul_science_2015}  and the driving term
$\cos(\omega t)\delta H_{\rm cot.}$, leads to the same consequence: a
mixing of the low energy states $|a\rangle$ and $|b\rangle$, and thus, to a larger Rabi
flop-rate.  This is illustrated in Fig.~\ref{fig2}b) where we show how
$\Omega$ changes with a transversal magnetic field. From the experimental
point of view,~\cite{\expESR} the static transversal field $B_x$ is also
required in order to have a finite tip polarization.

\section{Discussion}
We have analyzed the effect of an applied radiofrequency bias voltage on
the DC tunneling current through a magnetic adatom. Our basic assumption is
that this driving voltage leads to a modulation of the tunnel junction
transmission with the time-dependent external electric field.
In other words, the hopping tunneling amplitudes are modulated, giving place to an off-diagonal  time-dependent term in the adatom's Hamiltonian, which takes the form of the Rabi flop rate. 

The amplitude of the modulation was estimated using Bardeen
transfer Hamiltonian theory to describe the tunneling
current.~\cite{Bardeen_prl_1961} Thus, we have approximated the
 potential barrier between the two electrodes, tip and metallic
surface, by a square potential. Although the potential in a real
STM junction clearly differs from this simple picture, it still can
provide results in quantitative agreement.~\cite{Lounis_arXiv_2014}
A clear improvement over this simple description would be in the form
of the Tersoff-Hamann  description of tunnel between a surface and a
probe tip.~\cite{Tersoff_Hamann_prb_1985} We should remark that the
square-potential
approximation only affects the hopping tunnel amplitudes from the
tip (or surface) to the magnetic adatom and vice versa. This analysis
already reveals an interesting consequence: the more opaque the
tunnel junction is, the larger the ESR signal is. This observation
is in consonance with the measurement of STM-ESR signals
on Ag (100) coated with a thin insulating layer of MgO,~\cite{\expESR} and it opens
the possibility of observing ESR signal on similar surfaces, such as
Cu$_2$N/Cu(100). Notice that, since the resonant signal is observed in
the tunneling current, a right balance between detectable tunnel currents and  opaque
character must be reached.

Our description of the tunneling process is based on
second-order perturbation theory, which is adequate to describe the behavior of magnetic
adsorbates deposited on a thin decoupling layer on top of the
metallic substrate, such as MgO on Ag(100),\cite{\expESR} or
Cu$_2$N.~\cite{Hirjibehedin_Lutz_Science_2006,Ternes_pss_2017,Delgado_Rossier_pss_2017}
The effect of the driving field is summarized in a single parameter:
the Rabi flop-rate $\Omega$. Thus, all our efforts have been oriented to
estimate $\Omega$. In doing so, we keep a second-order description of
the interaction of the quantum system (the adatom) with the
electronic baths (surface and tip electrons), using the Bloch-Redfield
approach to treat open quantum system.~\cite{Breuer_Petruccione_book_2002}
In addition, we assume that the small and fast-oscillating driving field
does not modified the dissipative dynamics.~\cite{Cohen_Grynberg_book_1998}
In this description, the variation of the tunneling amplitudes induced
by the radiofrequency potential leads to an oscillating perturbation
of the adatom Hamiltonian. This time-dependent contribution mixes the
stationary states of the adatom, giving place to a finite Rabi flop-rate.

We have applied our theory on three different models to simulate the STM-ESR mechanism. In
first place, to a single-orbital Anderson model, 
which reveals  that isotropic $S=1/2$ systems
can be ESR active with a Rabi flop rate proportional to the tip polarization $P_T$.
A generalization of this model corresponds to a single-orbital multispin system with $S>1/2$, which can also include magnetic anisotropy.
Our analysis showed that the resulting Rabi frequency is proportional to the matrix elements $\langle a|\hat S_z|b\rangle$ of the spin operator between the two states connected by the resonant signal. This finding is in agreement with the observation of similar Rabi flop-rates for different atomic species.~\cite{\expESR}  
 Thus, the proposed mechanism does not rely on
a particular symmetry of the adsorbed adatom, neither on the adatom
magnetic anisotropy or total spin. This ubiquity is in agreement with
the experimental observation of STM-ESR for a variety of adatoms adsorbed
on MgO,~\cite{\expESR} including the Ti-H complex behaving as a $S=1/2$
spin.~\cite{Willke_Bae_science_2018,Willke_Yang_arXiv_2018,Yang_Bae_prl_2018}
This should work also with other spin-$1/2$ systems, including molecules
with $S=1/2$ spin centers like the Cu phthalocyanine. A similar analysis should also work for half-integer spins with strong hard-axis anisotropy, for which the ground state doublet satisfy $|\langle S_z\rangle|\approx 1/2$.

The resulting Rabi flop-rate depends on off-diagonal
 matrix elements mixing the two states connected by the ESR, and
 thus, it can be described as an effective AC magnetic field $B^{\rm
 eff}_{\perp}=2\hbar\Omega/g\mu_B$, whose orientation is parallel
 to the
 tip-polarization.  This is similar to the usual ESR where the AC field is perpendicular
to the field creating the Zeeman splitting,  and explains the need of 
an in-plane magnetic field in the experiments of Baumann {\it et al.}~\cite{Baumann_Paul_science_2015}

One prediction of the  single-orbital models is that
$\Omega$ is directly proportional to the tip polarization. 
This leads to a null contribution of the spin-unpolarized surface to the Rabi flop rate,
which in turn leads to
a linear dependence on current.  Although current dependences for
$S=1/2$ systems have not been reported to the best of our knowledge,
this result is in contrast with the observation of a current-independent
Rabi flop rate for Fe/MgO.~\cite{Willke_Paul_sciadv_2018} 
Hence, we employed a more sophisticated description of the adatom in terms of a
multiorbital Anderson Hamiltonian derived from a multiplet calculation. Using the well known Fe/MgO system as example, this model already pointed to an important result: when the orbital degrees of freedom of the adatom are accounted for, the modulation of the tunnel barrier by the AC electric field generates a finite Rabi flop rate even in the absence of current polarization. Due to the usually dominant contribution of scattering with surface electrons, the contribution associated to the surface overshadow the (current-dependent) tip part. This result is thus in agreement with the observed weak current dependence.~\cite{Willke_Paul_sciadv_2018}

%
Finally, although we have provided a way to calculate the intensity of
the driving term $\Omega$, a quantitative description is challenging:
it involves a precise knowledge of the hybridization functions with
the surface and tip, together with the density of states.  Here we
have used a rough estimation based on a flat-band model. Despite
the limitations, it allows us to get Rabi flop-rates high enough
to explain the observation of ESR signals, but the values are off the experimental
ones by a factor 10-20.~\cite{Willke_Paul_sciadv_2018} Here two
alternative improvements can be envisaged. On one hand, to use the
hybridization functions $V_{k;\ell}$ from a Wannier representation
of the DFT results, together with the PDOS $\rho(\epsilon)$ on the
whole energy interval. This approach is still problematic
due to both, slow convergence of the hybridization in the Wannier
bases, and the already poor DFT representation of the surface
hybridization.~\cite{korytar}  On the
other, one could use the flat band approach with constant hybridizations
using the cut-off as a fitting parameter. In both cases one finds an
additional problem: the most notable contribution is associated to the
spin-polarized tip, whose microscopic structure is basically unknown.

Our theory also predicts a finite Rabi flop-rate for spin-unpolarized
tunneling currents but, since the detection mechanism is based on
magnetoresistance, the DC current is in this case immune to the
radiofrequency, in accordance with the experimental observation.

\section{Conclusions}
In this work, we have demonstrated that the all-electrical spin resonance
phenomenon can be understood by the modulation of the tunnel junction
transmission with the time-dependent external electric field.  This,
in turns, originates an oscillating driving term on the adatom energy,
which can be understood as an effective magnetic field connecting the
two states involved in the ESR transition.

Our description is based on a perturbative treatment of the interaction
between the adatom and surface and probe tip. In particular, the electric
driving field leads to a perturbation of the adatom Hamiltonian,
which can be interpreted as an effective transversal magnetic field
coupling the eigenstates of the adatom.

%
The proposed mechanism leads to a Rabi
flop-rate barely dependent on the tunneling
current, in agreement with the experimental observations for
Fe/MgO.~\cite{Willke_Paul_sciadv_2018} 
Likewise, it predicts a linear dependence with the atomic spin component $\hat S_z$, in good agreement with the observation of STM-ESR weakly dependent on the atomic species.~\cite{\expESR}
In addition, it permits us to understand the interplay between the
tip-polarization, the magnetic anisotropy and the transversal
magnetic field. 

Future work involves improving the determination of the hybridizations in order to yield quantitative predictions.
Albeit the mentioned problems to calculate the Rabi flop-rate, an
accurate description of the ESR-lineshape also involves the relaxation
$T_1$ and decoherence $T_2$ times of the atomic spin. From a theory
point of view, a parameter-free description of the ratio $T_1/T_2$ is
really demanding. Both quantities have in general a completely different
dependence on the adatom magnetic anisotropy, longitudinal and transverse
magnetic fields.~\cite{Delgado_Rossier_pss_2017} Thus, even if we were able to reproduce
the excitation spectrum with an uncertainty smaller than $k_BT$, the
uncertainty in $T_1/T_2$ would be of several orders of magnitude.  As a
matter of fact, the extreme energy resolution of STM-ESR together with
the time resolution of STM pump-probe techniques
can be used to test the different theoretical methods.

\acknowledgments
We are grateful for many instructive discussions with T. Choi, J. Fern\'andez-Rossier, A. J. Heinrich,
C. Lutz and P. Willke.
NL, FD and JR acknowledge funding
from the Ministerio de Ciencia e Innovaci\'on grant MAT2015-66888-C3-2-R and FEDER funds. CW acknowledges support from Institute for Basic Science under IBS-R027-D1. FD acknowledges financial support from Basque Government, grant IT986-16 and Canary Islands program {\it Viera y Clavijo}  (Ref. 2017/0000231).

\appendix

\section{Review of the cotunneling theory \label{AppendixC}}
The tunneling Hamiltonian that changes the number of electrons in the correlated quantum system by one unit can be written as
\begin{equation}
{\cal H}_{tun}=\sum_{\alpha,\mathbf{i}}\left[V_{\alpha,\mathbf{i}}(t)f_{\alpha}^{\dagger}d_{\mathbf{i}}+V^{*}_{\alpha,\mathbf{i}}(t)d^{\dagger}_{\mathbf{i}}f_{\alpha} \right]
\equiv{\cal V}^- (t)+{\cal V}^+ (t)
\label{EqA1}
\end{equation} 
where $d^{\dagger}_{\mathbf{i}}$ creates an electron with quantum numbers $\mathbf{i}=(\ell, \sigma)$ with $\ell$ the orbital number and $\sigma=\sigma_{\alpha}$ the spin. Using second-order perturbation theory we can write an effective Hamiltonian acting only on the $N_0$-charge space, which hereafter we shall refer as {\it neutral charge state}. If we denote by $|M_\pm\rangle$ the eigenstates of the decoupled electrode+central region with $N_0\pm 1$ electron, we can write the matrix elements between states $|N\rangle$ and $|N'\rangle$ as
\begin{equation}
\sum_{M_-}\frac{{\cal V}^+ |M_-\rangle\langle M_-|{\cal V}^-}{E_{M_-}-E_0}
+\sum_{M_+}\frac{{\cal V}^- |M_+\rangle\langle M_+|{\cal V}^+}{E_{M_+}-E_0},
\label{Heff_cot}
\end{equation}
where $E_0$ is the ground state energy of the (decoupled) system with $N_0$ electrons in the central region.
Roughly speaking, the cotunneling approach will remain valid as long as 
\begin{equation}
\left|\frac{\langle N|{\cal V}^{\pm}|M_{\mp}\rangle}{E_{M_{\mp}}-E_N}  \right|\ll 1
\end{equation}
Now we have to evaluate the corresponding matrix elements.
First, let us consider the $|N\rangle$ states can be written as $|N\rangle =|n\rangle \otimes | \Psi\rangle$ where $| \Psi\rangle$ is a multi electronic Slater determinant that describes independent Fermi seas of left and right electrodes. They describe an arbitrary state of the central island and states with an electron-hole pair in the electrodes. For the $N_0\pm 1$ electron states, we write $|M_{\pm}\rangle =|m_{\pm}\rangle \otimes | \Psi_\mp\rangle$, where now $|\Psi_\mp\rangle$ is a Slater state for the electrodes with one electron more (+) or less (-) than the $N_0$ manifold. 

If we denote by $|0\rangle$ the ground state of the electrodes in the Fermi sea with no excitations in the neutral charge state, we can write $|\Psi\rangle\equiv f_{\alpha}^{\dagger}f_{\alpha}|0\rangle$, where we are creating an electron-hole pair with quantum number $\alpha$. For the states with one electron excess (defect) we will have $|\Psi_-\rangle=f_{\beta}f_{\alpha}^{\dagger}f_{\alpha}|0\rangle$ and $|\Psi_+\rangle=f_{\beta'}^\dag f_{\alpha}^{\dagger}f_{\alpha}|0\rangle$. The zero-temperature occupation of an electrode state $\alpha$ is then given by $n_\alpha=\langle \Psi|f_{\alpha}^{\dagger}f_{\alpha}|\Psi\rangle$, which can only take the values 0 or 1 for electrons.

The matrix element of the electrode operator in Eq. (\ref{Heff_cot}) selects only one term in the electrode part of the sums $\sum_{M_\pm} =\sum_{m_\pm}\sum_{\Psi_\mp}$. Then one can write
\beqa
&&\sum_{\Psi_+}\langle\Psi|f_{\gamma}| \Psi_+\rangle=(1-n_\gamma)\delta_{\beta \gamma},
\\
&&
\sum_{\Psi_-}\langle\Psi|f_{\gamma}^\dag| \Psi_-\rangle=n_\gamma\delta_{\beta \gamma}.
\eeqa
Making the corresponding substitution into Eq. (\ref{EqA1}) we get
%
\beqa
\sum_{M_-}\frac{\langle N|{\cal V}^+ |M_-\rangle\langle M_-|{\cal V}^-| N'\rangle}{E_{M_-}-E_0}&=&
\sum_{\alpha\alpha'}
\left( 1-n_\alpha\right) 
\crcr
&&\hspace{-3.cm}\times \langle n|\hat{T}_-(\alpha\alpha';t)|n'\rangle
\langle \Psi|f_{\alpha}f_{\alpha'}^{\dagger}|\Psi\rangle ,
\label{Hquasi1}
\eeqa
and
\beqa
\sum_{M_+}\frac{\langle N|{\cal V}^- |M_+\rangle\langle M_+|{\cal V}^+| N'\rangle}{E_{M_+}-E_0}&=&
\sum_{\alpha\alpha'}n_\alpha
\crcr
&&\hspace{-3.cm}\langle n|\hat{T}_{+}(\alpha\alpha';t)| n'\rangle
\langle \Psi|f_{\alpha}^{\dagger}f_{\alpha'}|\Psi\rangle,
\label{Hquasi2}
\eeqa
where we have introduced the transition amplitude operators $\hat{T}_\pm(\alpha\alpha';t)$, whose matrix elements are given by
\beqa
\langle n|\hat{T}_-(\alpha\alpha';t)|n'\rangle &=&  
\sum_{m_-,\ell\ell'}\frac{V^*_{\alpha,\ell}(t)V_{\alpha',\ell'}(t) }{E_{m_-}-E_0+\epsilon_{\alpha} }
\gamma_{nn'}^{m_-}(\alpha \ell,\alpha' \ell')
\crcr &&
\label{Tminus}
\\
\langle n|\hat{T}_+(\alpha\alpha';t)|n'\rangle &=&  
\sum_{m_+,\ell\ell'}\frac{V_{\alpha,\ell}(t)V^*_{\alpha',\ell'}(t) }{E_{m_+}-E_0-\epsilon_{\alpha}  } 
\gamma_{nn'}^{m_+}(\alpha \ell,\alpha' \ell').
 \crcr &&
\label{Tplus}
\eeqa
Equations (\ref{Hquasi1})-(\ref{Hquasi2}) can be simplified
by taking into account that $\sum_{\alpha,\alpha'}n_\alpha
\langle \Psi|f_{\alpha}f_{\alpha'}^{\dagger}|\Psi\rangle= \langle
\Psi|\sum_{\alpha\alpha'}f_{\alpha}f_{\alpha'}^{\dagger}|\Psi\rangle$.
Hence, when we restrict the reservoir states to single electron-hole
pairs $|\Psi\rangle$, we recover Eq. (\ref{Hcotun_eff}) of the main text.

Notice that here, the central region is described by a time-independent
Hamiltonian ${\cal H}_C$ and thus, stationary eigenvalues and eigenvectors
can be introduced without loss of generality. If, on the other hand,
the time dependence is contained in  ${\cal H}_C$, the stationary
description becomes ill defined.  Despite this issue, a similar analysis can
be carried out provided the driving field is small enough. For instance,
if the central region is under the effect of a time-dependent electric
potential $|V_{ac}| \ll |E_{m_\pm}-E_0|$, we can still work in the
pseudo-stationary states $|N\rangle$ and $|M_\pm\rangle$, while the
energy differences becomes
\beq
E_{m_\pm}(t)-E_0(t)\approx E_{m_\pm}-E_0\mp V_{ac}\cos(\omega t).
\eeq

\section{Energy integrals ${\cal I}^\pm(z_1,Ec,z_2)$ \label{AppenI}}
The following energy integrals can be done analytically by deformation in the complex energy plane
\beq
{\cal I}^\pm(\epsilon_0,E_c,\mu)={\cal P}\int_{-E_c}^{E_c}d\epsilon\frac{f^\pm_\mu (\epsilon)}{\epsilon_0-\epsilon}.
\eeq
The results are given by
\beqa
{\cal I}^+(\epsilon_0,E_c,\mu)\approx &&
\ln (2 \pi ) -\ln ( E_c-\mu +\epsilon_0)
\crcr
&&\hspace{-1.cm}+
{\rm Re}\left[ \psi ^{(0)}\left(\frac{1}{2}-\frac{i   (\epsilon_0-\mu )}{2
   \pi }\right)\right] , 
   \label{DefIp}
\eeqa
and
\beqa
{\cal I}^-(\epsilon_0,E_c,\mu)\approx \ln\left|\frac{\epsilon_0-E_c}{\epsilon_0+E_c}\right|-{\cal I}^+(\epsilon_0,E_c,\mu).
  \label{DefIm}
\eeqa
where $\psi^{(0)}(x)$ is the digamma function and the arguments satisfy
$-\epsilon_0+\mu< E_c<\epsilon_0-\mu$. The above approximations correspond
to asymptotic expansions for $E_c\gg 1$. Notice that here we have used
a description in terms of dimensionless variables, which is equivalent
to measure all energies in units of $k_BT$. In the case of interest,
$E_c\gg 1$ and $E_c/\epsilon_0\sim 1$. In this limit, to lowest order
in $1/E_c$, we have that
\beq
{\cal I}^\pm(\epsilon_0,E_c,\mu)\sim \mp \ln\left(\frac{1}{1+\epsilon_0/E_c}\right).
\eeq




\end{document}